# Trust in Construction AI-Powered Collaborative Robots: A Qualitative Empirical Analysis

Newsha Emaminejad[1] and Reza Akhavian, Ph.D., M. ASCE[2]

[1]Graduate Student, Dept. of Civil, Construction, and Environmental Engineering, San Diego State University, San Diego, CA. Email: nemaminejad8591@sdsu.edu
[2]Associate Professor, Dept. of Civil Engineering, San Diego State University, San Diego, CA (corresponding author). ORCID: https://orcid.org/0000-0001-9691-8016. Email: rakhavian@sdsu.edu

## ABSTRACT

Construction technology researchers and forward-thinking companies are experimenting with collaborative robots (aka cobots), powered by artificial intelligence (AI), to explore various automation scenarios as part of the digital transformation of the industry. Intelligent cobots are expected to be the dominant type of robots in the future of work in construction. However, the black-box nature of AI-powered cobots and unknown technical and psychological aspects of introducing them to job sites are precursors to trust challenges. By analyzing the results of semi-structured interviews with construction practitioners using grounded theory, this paper investigates the characteristics of trustworthy AI-powered cobots in construction. The study found that while the key trust factors identified in a systematic literature review -conducted previously by the authors- resonated with the field experts and end users, other factors such as financial considerations and the uncertainty associated with change were also significant barriers against trusting AI-powered cobots in construction.

## INTRODUCTION

The construction industry continues to adopt technologies that help address its grand challenges such as poor safety and productivity records and shortage of skilled labor. Collaborative robots (aka cobots), is a prime example of such technologies, and is increasingly becoming a major component of this evolution (Afsari et al. 2018). Cobots can revolutionize the construction industry by making tedious, repetitive, and physically demanding tasks safer, more efficient, and with higher cost effectiveness (Follini et al. 2020). They are equipped with advanced sensors and safety features that allow them to perform tasks with precision and avoid accidents. Cobots, are now being used in a wide range of construction tasks, from bricklaying to welding, 3D printing, heavy lifting, manual material handling, and inspection (Burden, Caldwell, and Guertler 2022). By augmenting human workers, cobots help reduce fatigue and increase productivity, while also freeing up workers to focus on more complex tasks (Ma, Mao, and Liu 2022). The use of cobots in construction also helps improve project timelines and reduce overall costs, making them an attractive investment for companies looking to stay competitive in the ever-evolving construction industry (Veloso et al. 2012). However, despite the many benefits of cobots, there are also major challenges that need to be addressed before they can be fully integrated into the construction jobsites. One of the most important of these is building trust between construction workers and cobots (Calitz, Poisat, and Cullen 2017). Trust is a complex concept that can be defined as the belief in the reliability and integrity of someone or something (Jian, Bisantz, and Drury 2000). In the context of construction, trust between workers and cobots is important for ensuring that these technologies are used effectively and safely (Emaminejad, Maria North, and Akhavian 2021).

**RESEARCH BACKGROUND**

Recent studies have focused on the acceptance and trust of collaborative robots (cobots) in industrial workplaces. These studies have examined factors such as socio-technical systems, interpretability, predictability, transparency, reliability, framing, and human factors. One study proposed a conceptual model that combined the Unified Theory of Acceptance and Use of Technology (UTAUT) and Socio-Technical Systems theory (STS) to understand critical factors influencing the acceptance of cobots and drive perceived work performance improvement at the organizational level (Prassida and Asfari 2022). Another study explored the effectiveness of different light- and motion-based cobot signals in various collaborative mini-games. The studies recommend design improvements for cobots, including programming and interface designs, educational technologies, and careful selection of information to counteract negative effects of failures (Mara et al. 2021). In a study by Michaelis et al. (2020), interviews with manufacturing experts revealed that design improvements for cobots, including programming and interface designs, and educational technologies are required to support collaborative use (Michaelis et al. 2020). Another study by Kluy and Roesler et al. (2021) investigated the influence of transparency and reliability on perception of and trust towards cobots (Kluy and Roesler 2021). Kopp et al. (2022) explored how framing and perceived cooperativeness affect anthropomorphization and human-robot trust in inexperienced factory workers. Lambrechts et al. (2021) emphasized the need for phased implementation and the leadership role in cobot success by highlighting the importance of linking human factors to the future of work and focuses on reskilling and upskilling logistics professionals in response to robotization. The authors conducted a literature review on trust in cobots within construction projects and identified four trust dimensions: transparency and interoperability, reliability and safety, performance and robustness, and privacy and security. Trust across these dimensions is crucial for workers to embrace and collaborate effectively with cobots. Lack of trust can impede cobot adoption and implementation in construction. Trust is essential for collaboration, successful implementation, and worker comfort with cobots (Emaminejad and Akhavian 2022). The existing literature primarily focuses on cobots in manufacturing or other industries, with limited research on how these cobots are perceived in the construction industry. The current study aims to fill this gap by obtaining insights from practitioners in the AEC industry.

**METHODOLOGY**

The previous literature exploration by the authors has resulted in identifying the technical and psychological factors that increase trust in cobots at both the organizational and individual levels. The authors then formulated the constructs of a theoretical model called Construction Robotics Adoption Drivers (ConRAD) by interviewing construction practitioners, with the goal of confirming or refuting previous findings from the literature review and gaining practical insights. The scope of the interviews was also covered other aspects such as the needs and challenges in training and upskilling construction practitioners for digital transformation that includes AI-powered cobots. The grounded theory was utilized as a research tool for systematic analysis of data to develop a theory that is grounded in the participants' experiences and perspectives (Soliman and Kan 2004). Figure 1 shows how grounded theory was applied in this study (Di Gregorio 2003).

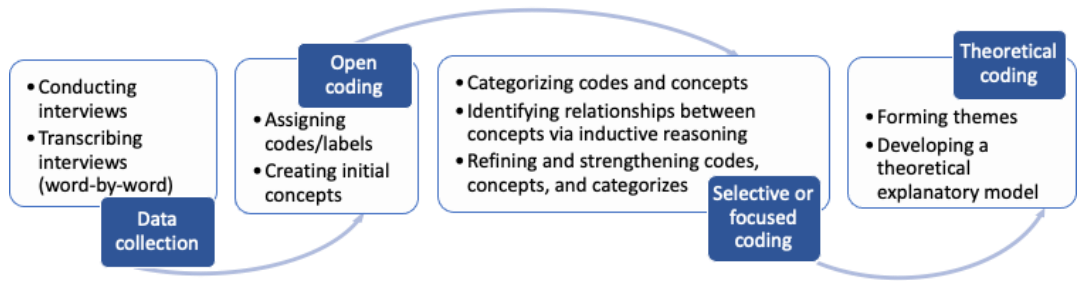

Figure 1. Schematic overview of qualitative data analysis using grounded theory

Table 1. Interview questions.

| |
|---|
| 1. What is your opinion about using robots in construction? |
| 2. [after watching a short video involving three different types of cobots in construction settings] Based on what you just watched, what is your opinion about using intelligent cobots in construction projects? |
| 3. In your opinion, what are the challenges limiting the adoption of intelligent cobots in construction? |
| 4. What makes you (not) trust an intelligent cobot? what aspect(s) (e.g., technical, management, financial, psychological, etc.) have the most impact on your opinion about trusting them? |
| 5. In your opinion, is trusting intelligent cobots a top-down approach or bottom-up in construction projects and company organizational structure? Meaning that should it first be trusted by workers than the managers or the other way around? |
| 6. To prepare project teams to work with intelligent cobots, what types of training (e.g., technical, soft skills, social) would you like to see for: either of these groups:<br>    A. Field personnel (e.g., workers, foremen, superintendents)<br>    B. Office personnel (e.g., project engineers, project managers, project executives) |
| 7. A. Do you <u>think</u> that intelligent robots <u>are poised to</u> replace workers, so workers are assigned to higher-level roles? If so, in what capacity? |
| 7. B. Do you <u>want</u> to see intelligent robots <u>will</u> replace workers, so workers are assigned to higher level roles? If so, in what capacity? |
| 8. We will soon start a nationwide survey on this topic. Is there anything you are curious about and would like to gauge the opinion of the construction industry? |

The research team conducted in-depth interviews (approved by the San Diego State University Institutional Review Board (IRB)) with 11 construction professionals who met specific criteria, including having experience working in the AEC industry, working in various company types and construction industry sectors, having experience working with technology in the AEC, and conducting research on topics related to the implementation of intelligent cobots/robots in the AEC. The sample included a diverse group of participants including two Presidents/CEOs, an Executive Vice President, a Senior Director, a Robotics Lead, three VDC Managers, and three Superintendents from five different construction companies in the US. The interviews were conducted individually and online via Zoom, using a pre-approved protocol that included a description of the process, verbal consent, and a set of 8 semi-structured questions (Table 1). However, free discussion was allowed to unfold in a more natural conversation with unscripted questions being added in order to gain a clearer understanding of emerging concepts. The interviewees were shown a video of collaborative robots in practice, and a definitions table was provided to ensure a clear understanding of terms used during the interview. They took place between May and August 2022 and resulted in a total of 11.5 hours of audio-recorded conversations, which were transcribed word-for-word for accuracy. NVivo 12 software was used to code the data collected during the interviews into themes and categories and theories were continuously revised and refined to ensure that they remained grounded in the participants' experiences and perspectives.

**RESULTS AND DISCUSSION**

The main concepts and themes emerging from the interviewees' perceptions are summarized in Figure 2, which will be discussed in more detail in the following sections.

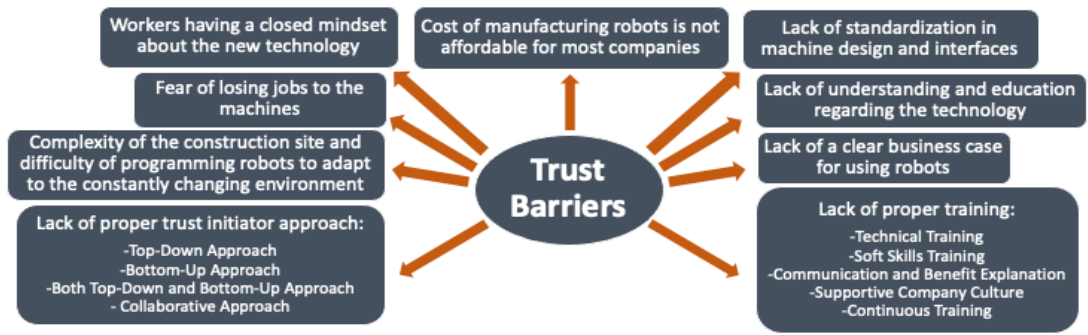

**Figure 2. Main trust barriers based on experts' perceptions.**

**General knowledge and opinion**. To begin collecting information through interviews in this research, it is important to assess the interviewee's level of general knowledge about robotics role, and impression of working alongside cobots in construction projects. Inadequate familiarity and basic understandings of a technology can pose a significant obstacle in building trust and gaining acceptance of that technology. Most of the interviewees had positive opinions about the potential of robotics in construction and there is a general consensus that robots have the potential to revolutionize the construction industry by increasing safety and productivity and facilitating collaboration between different parties. One of the interviewees referred to the use of cobots as a great equalizer between automation and human supervision for ensuring consistent product quality. Many of the interviewees confirmed an original hypothesis of this research that the robots can be used to perform repetitive or hazardous tasks, such as lifting heavy items, reducing the risk of back and soft tissue injuries, and freeing up human workers to do more complex work. However, the level of optimism regarding the current state of robotics in construction varied among the interviewees. Some interviewees have not seen any robots in-use and believed that they are only used for presentations and showcases. There was a sense that the construction industry is still in the early stages of exploring this technology, and there will be further advancements in the future to expect. Some interviewees expressed concerns about the effectiveness of robots in the construction industry. It was noted that cobots may not be suitable for all tasks, as not all jobs are repetitive, and they may require significant time and monetary invest also emphasized that robots should be able to adapt to changing conditions on construction sites and be able to learn from their experiences. Moreover, regarding the types of projects, some of them believed that large projects like bridge construction may be able to absorb the cost and benefits from using cobots, while smaller projects may struggle to justify the expense. Two superintendents raised concerns about safety, mentioning that they have not encountered robots at their workplace and view the idea as being in the testing phase. There were also concerns about the impact of robotics on human workers and their job security, which alludes the need for providing appropriate training and education to adapt to the changing nature of the construction industry. However, some other interviewees mentioned that robots can free up personnel to do more mentally intensive tasks that require creativity and complex decision-making, and that AI can be used to streamline workflows.

**Potential challenges for adoption**. One of the primary challenges for the adoption of intelligent cobots in the construction industry is the cost of manufacturing robots. In addition, standardization is essential to the implementation of robots in construction, as there are many engineering firms, and each project is configured differently. Different manufacturers have different designs and interfaces for their machines, which makes it difficult to develop standard training for workers to use them. The lack of understanding and education regarding the technology is another significant challenge. Construction companies are not aware of the benefits of using robots and new technologies. The implementation of cobots requires education and information campaigns to raise awareness and convince construction companies to use robots. The lack of a clear business case for using cobots is also an obstacle to the adoption of them in the construction industry. Companies need to see a clear return on investment before they invest in the technology. The complexity of the construction site and the difficulty of programming robots to adapt to the constantly changing environment, which is very challenging even for humans are also major challenges. Construction sites are often dynamic and chaotic, and there are many variables that can affect the ongoing work. Cobots need to be programmed to handle unpredictability, which requires sophisticated technology and algorithms. Moreover, potential closed-minded mentality of workers and the fear of losing their jobs to the machines are also big challenges. People tend to have a closed mindset about the benefits of a new technology and prefer to do things the old-fashioned way. People who come from a demographic that primarily depends on physical labor to make a living have a greater fear of losing their jobs to machines. They'll eventually get comfortable with the machines, but the biggest challenge lies in getting them to embrace the change.

**Trust barriers**. Many interviewees mentioned the importance of cobot's ability to perform tasks efficiently, without errors, and within a reasonable amount of time. Demonstrating cobot's ability to adapt to different environments, and process information can also help build trust. The interviewees also emphasized the need for proof of concept and testing to identify practical use-cases for the cobot and to find its limitations. Close monitoring and quality control can help mitigate safety concerns, and the use of case studies and sales pitches to demonstrate successful implementations of the technology in other organizations can help build trust. Moreover, several interviewees raised concerns about the "fear of the unknown" and the lack of understanding of how the technology works. Interviewees emphasized the importance of transparency and building trust with clients and stakeholders. They suggested that manufacturers must pay attention to quality control to avoid errors that could result in recalls or retraining. They also pointed out that multiple masons on a job site can provide more quality control than a single automated system and how the robot manufacturer justifies its ability to control quality should be clarified. The technical skills of the individual paired with the cobot are also essential. The aforementioned points align with the existing literature and validate the fact that trust between humans and robots is heavily influenced by factors such as reliability, performance, transparency, and interpretability. Some interviewees also highlighted the importance of the size and appearance of the cobot, suggesting that smaller cobots may be less intimidating and more approachable than larger humanoid robots. Interviewees also emphasized the importance of safety and control, with several suggesting the need for manual human interfaces to override the cobot when necessary. Having control over the cobot is essential to gain workers' trust. A notable observation was that only three interviewees expressed worries about privacy and security when they were asked, specifically with the use of cameras in cobots, and suggested that manufacturers should implement appropriate measures to mitigate these concerns. They acknowledged that these concerns may be more prevalent among companies involved in larger, public construction projects.

**Trust initiation approach**. There was no clear consensus among participants on whether trusting intelligent cobots is a top-down or bottom-up approach in construction projects and company organizational structure. Some interviewees suggested that a top-down approach is essential since it is the management who makes decisions about investing in technology and providing training and support to workers. Others suggested a bottom-up approach is necessary since the workers are the ones who will use the technology and must feel comfortable and familiar with it. They need to be trained and allowed to ask questions and provide feedback. However, a majority of the interviewees suggested the need to a blend of the top-down and bottom-up approaches to build trust in intelligent cobots. They believed that managers should educate the workers about the benefits of cobots, involve them in the decision-making process, provide training, and allow them to provide feedback to improve the technology's performance. At the same time, top management should investigate the technology, research its benefits, and set expectations for the rest of the organization to embrace it. Some interviewees, on the other hand, suggested that depending on the situation, building trust in intelligent cobots may require a collaborative approach, where the company invests in technology, provides training, and communicates its benefits to workers who should show willingness to embrace the technology and provide feedback to improve it. Ideally, trust in cobots will only be achieved when both the management and workers are convinced of its benefits and potential to enhance efficiency and safety.

**Training**. Analyzing the responses regarding the types of training required to prepare project teams to work with intelligent cobots, several common themes emerged but overall, the responses suggest that there is no one-size-fits-all solution to preparing project teams to work with intelligent cobots. First, there was a consensus among the interviewees that technical training is crucial when working with robots, and safety should be a top priority. This includes training on how to use and operate the robots, as well as how to use them effectively to improve productivity. Interviewees also suggested that training should be tailored to the needs of the organization, and management needs to be involved in the process to ensure that everyone is on board with the change. Second, soft skills training may be necessary to ensure that personnel are open to change and willing to adapt to new technology. Interviewees suggested that this could include training in communication, teamwork, and adaptability. Third, interviewees suggested that it is important to communicate the benefits of the technology in a way that is easily understandable to different groups, such as the field and office personnel both. This could involve creating specialized roles focused on using the new technology and moving from job site to job site and ensuring that communication channels are established between different levels of personnel. Fourth, it is important to have a supportive organization culture where office personnel support the field personnel in adopting to the new teamwork schemes. Change agents could be identified among superintendents who can demonstrate to their colleagues the benefits of the technology, and education, training, and experience should be in balance for the best results. Fifth, the interviewees suggested that training should be a continuous process, and it is important to have hands-on training in an environment similar to the construction site to build trust.

**Job security**. It appears that the prevailing view from interviewees is that while intelligent cobots are set to replace workers in specific roles, they will not completely supplant the labor force. Rather, they will transform job characteristics and necessitate different skill sets for job execution. They suggest that the integration of cobots will change the nature of jobs, allowing workers to focus on more skilled and higher-paying tasks. This shift will lead to more efficient and effective workflows that will allow companies to expand and create more jobs. Several interviewees believed that the use of cobots in the workforce is a natural progression given the trajectory of technology. The key challenge for the construction industry will be to find the right balance between the use of cobots and employing humans to perform tasks that require creativity and complex decision-making, and to ensure that workers are comfortable and confident in working alongside cobots and supervise them. The majority of them believed that cobots will not replace human workers, but instead augment their work which will help with the labor shortage. They suggested that industries will require workers with skills such as programming the robots, and new interdisciplinary jobs will be created as a result. Some believe that workers will be needed to perform certain tasks that cobots are not flexible enough to handle, while others suggest that workers will focus on developing skills in areas where machines are not efficient, such as creativity and emotional intelligence. They mentioned that workers' roles will shift to quality control and fine-tuning of the work done by cobots, as well as maintenance and repair of cobots. They also believed that workers with high-level expertise, such as welders and masons, working with technology developers, are vital to improve the cobots performance.

## CONCLUSION

Through interviews with 11 experts in construction industry, this study tries to understand factors that influence trust in cobots and adoption barriers in the construction industry. This understanding is critical for successful implementation and adoption of cobots and new technologies that are trusted and accepted by workers. The overall consensus was that the use of cobots has the potential to revolutionize the construction industry by increasing safety, productivity, and collaboration between different parties. However, there were some significant hurdles that need to be overcome, including the high cost of manufacturing robots, lack of standardization, complexity of the construction site, and fear of job loss. Additionally, building trust between humans and robots requires a multi-faceted approach that addresses a wide range of concerns, from technical functionality to ethical implications. Therefore, it is necessary to have a thoughtful and well-planned approach that includes education and awareness-raising campaigns to convince construction companies to use cobots and make a clear business case for investment. Technical training, soft skills training, effective communication, and a supportive company culture are also necessary for this approach. Furthermore, finding the right balance between the use of technology and the employment of human workers is essential.